\documentstyle[12pt]{article}
\begin{document}
\vspace*{-3.5cm}
\hfill{\small KL--TH 96/14}\\[4mm]
\begin{center}
{\Large\bf Quantum Tunneling of Spin Particles\\
in Periodic Potentials with Asymmetric Twin Barriers}
\\[12mm]
{\large J.--Q. Liang}\\
{\it Department of Physics, University of Kaiserslautern, D--67653\\
Kaiserslautern, Germany,\\
Institute of Theoretical Physics, Shanxi University, Taiyuan, Shanxi 030006, P.R. China,\\
Institute of Physics, Chinese Academy of Sciences, Beijing 100080, P.R. China}\\[4mm]
{\large H.J.W. M\"uller--Kirsten, Jian--Ge Zhou and F. Zimmerschied}\\
{\it Department of Physics, University of Kaiserslautern, D--67653\\ Kaiserslautern,
Germany}\\[4mm]
{\large F.-C. Pu}\\
{\it Department of Physics, Guangzhou Teachers College, Guangzhou 510400, P.R. China,\\
Institute of Physics, Chinese Academy of Sciences, Beijing 100080, P.R. China}\\[1cm]
{\bf Abstract}
\end{center}
The tunneling effect of a periodic potential with an asymmetric twin barrier per period is
calculated using the instanton method.  The model is derived from the Hamiltonian of a
small ferromagnetic particle in an external magnetic field using the spin--coherent--state
path integral.  The instantons in two neighbouring barriers differ and lead to different level
shifts $\triangle\epsilon_1, \triangle\epsilon_2$.  We derive with Bloch theory the
energy spectrum which has formally the structure of an energy band.  The spectrum depends
on both level shifts.  The removal of Kramer's degeneracy by an external magnetic field
is discussed. In addition we find a new kind of quenching of macroscopic quantum
coherence which is irrelevant to Kramer's degeneracy.

\vfill
\begin{center}
PACS numbers: 11.15.KC, 03.65.Db, 11.10.St, 73.40.GB
\end{center}
\newpage

\noindent
{\large \bf 1. Introduction}\newline\\
Quantum tunneling with various solvable potentials such as the double--well \cite{1},
inverted double--well \cite{2} and sine--Gordon \cite{3} potentials have been studied
extensively in the framework of the instanton method.  Instantons in field theory of
$1+0$ dimensions are viewed as pseudoparticles with trajectories existing in 
barriers, and are therefore responsible for tunneling.  Since instanton trajectories
have zero energy and nontrivial topological charge, and interpolate between 
degenerate vacua, the instanton method is only suitable
for the calculation of tunneling between neighbouring vacua.  Recently a new type of
pseudoparticle configuration has been constructed \cite{4,5} and dubbed
``periodic instanton''.  These periodic instantons have finite, nonzero energy
and are found to be useful in the calculation of tunneling at excited states \cite{6}.

In the case of the double--well potential the amplitude for transitions between
degenerate vacua induced by instantons results in the level shift of the
degenerate ground state.  In the case of the periodic potential the 
instanton induced transition amplitude is valid for tunneling through only
one of the barriers and is the same for all barriers.  The tunneling amplitude 
results in the level splitting which extends to an energy band \cite{3,6}
by translation symmetry, as follows from Bloch theory.  

The present paper is devoted to solving the tunneling problem with a periodically recurring 
asymmetric twin barrier which leads to two kinds of instantons.  This case is
of interest in its own and worthy of development of the appropriate instanton technique,
and is also of practical importance for macroscopic quantum tunneling \cite{7} in
magnetism \cite{8}.  

The layout of the paper is as follows.  In sections 2 and 3 the effective Lagrangian is
derived from the Hamiltonian of a small ferromagnetic particle in a
constant magnetic field.  The exact instanton solutions existing in the barriers are given.
We also calculate the total action along the instanton trajectories (the socalled
instanton mass).  Section 4 is devoted to the calculation of the energy spectrum with the
help of Bloch theory.  The resulting formula involves two level shifts to be 
determined from tunneling amplitudes.  In section 5 the tunneling amplitudes are 
obtained with the instanton method up to and including the one loop
correction.  Finally we discuss the quenching of macroscopic quantum coherence 
(MQC), i.e. that of the energy band due to the topological phases of
the effective wave functions of the small ferromagnetic particle in the
magnetic field.\newline\\

\noindent
{\large \bf 2.  The effective Lagrangian of a small ferromagnetic particle in a constant
magnetic field}\newline\\

We consider a giant spin in a constant magnetic field described by the Hamilton operator
\begin{equation}
{\hat H}= K_1{{\hat s}_z}^2 + K_2{{\hat s}_y}^2 - 2\mu_BB{{\hat s}_y}
\label{2.1}
\end{equation}
where ${\hat s}_i, i=x, y, z,$ are spin operators satisfying the usual commutation relation
$ [{\hat s}_i, {\hat s}_j] = i\epsilon_{ijk}{\hat s}_k $ (using natural units throughout), and
$\mu_B$ is the Bohr magneton.  The model describes in the absence of the
magnetic field ${\bf B}$ XOY easy plane anisotropy \cite{9} and an easy axis
along the x--direction with the parameters $K_1>K_2>0$.  The external magnetic
field is assumed to be perpendicular to the easy direction.  Our purpose in the
following is to convert the quantum spin system into a potential problem
suitable for the study of quantum tunneling.  To this end we start from the coherent
state representation of the matrix element of the time--evolution operator with 
the Hamiltonian of eq. (\ref{2.1}).  With the help of the spin--coherent 
path integral we obtain
\begin{equation}
<{\bf n}_f|e^{-2i{\hat H}T}|{\bf n}_i> = e^{-i(\phi_i-\phi_f)s}{\cal K}(\phi_f, t_f; \phi_i, t_i)
\label{2.2}
\end{equation}
where
\begin{equation}
{\cal K}(\phi_f, t_f;\phi_i, t_i) = \int{\cal D}\{\phi\}{\cal D}\{p\}e^{i\int^{t_f}_{t_i}{\cal L}
(\phi, p)dt}
\label{2.3}
\end{equation}
is the path integral in phase space with canonical variables $\phi$ and $p = s\cos\theta $.
Also
\begin{equation}
{\cal L} = \dot {\phi}p - H(\phi, p)
\label{2.4}
\end{equation}
is the phase space (or first order) Lagrangian.  The Hamiltonian
\begin{equation}
H = \frac {p^2}{2m(\phi)} + V(\phi)
\label{2.5}
\end{equation}
has position dependent mass $m(\phi)$ and potential
\begin{eqnarray}
m(\phi)& =& \frac {1}{2K_1(1-\lambda\sin^2\phi +\alpha\lambda\sin\phi)},\nonumber\\
V(\phi)&=& K_2s^2(\sin^2\phi-2\alpha\sin\phi +\alpha^2)
\label{2.6}
\end{eqnarray}
respectively.  The potential minima are chosen to be zero (see Fig.1).  
In these expressions
$$
\lambda\equiv \frac{K_2}{K_1}, \;\; \alpha\equiv\frac{\mu_BB}{sK_2}
$$
are dimensionless parameters whose values are correspondingly indicative of the
degree of anisotropy of the ferromagnetic particle and the asymmetry of the
twin--barrier (Fig.1).  The kets $|{\bf n}_i>$ and
$|{\bf n}_f>$ denote the initial and final spin--coherent-- states and $2T\equiv
t_f - t_i $ is the time difference.  Here ${\bf s}= s(\sin\theta \cos\phi, 
\sin\theta\sin\phi, \cos\theta)$ is visualised as a classical spin vector with
spin number $s$, polar angle $\theta $ and azimuthal angle $\phi $.
In the above derivation, the large spin limit $s>>1$ has been used since giant
spins with spin quantum number $s>>1$ are believed to correctly describe
ferromagnetic grains.  To simplify the model we also consider the tunneling
near the easy plane, i.e. for $\theta\approx\frac{\pi}{2}$. Our rule of simplification
is then that for $s\rightarrow \infty, \cos\theta \rightarrow 0$ the product
$s\cos\theta $ is to remain finite.  The phase factor $e^{-i(\phi_f - \phi_i)s}$
in the transition amplitude eq. (\ref{2.2}) is the characteristic of the spin
system and can be put into the Lagrangian, i.e. the expression 
$\phi_f - \phi_i = \int^{t_f}_{t_i} \dot{\phi} dt $,
and can be identified as a Wess--Zumino term \cite{9}. The effect of the phase factor
has been studied extensively \cite{8, 10}.  The position dependent
kinetic term requires that one should start from the phase space path integral \cite{11}, 
i.e. eq.(\ref{2.3}).  Integrating out the momentum in the path integral eq. (\ref{2.3}), 
we obtain the usual Feynman propagator in configuration space, i.e.
\begin{equation}
{\cal K}(\phi_f, t_f; \phi_i, t_i) = \int{\cal D}\{\phi\} e^{i\int^{t_f}_{t_i}
{\cal L}(\phi, \dot{\phi})dt}
\label{2.7}
\end{equation}
with the second order Lagrangian
\begin{equation}
{\cal L} = \frac {1}{2}m(\phi) {\dot{\phi}}^2 - V(\phi)
\label{2.8}
\end{equation}
\newline
\newpage

\noindent
{\large \bf 3. Exact instanton solutions, instanton ``mass'' and reduced tunneling model}\newline\\

The instanton solution in $1+0$ dimensions is a trajectory of the pseudoparticle
which exists in the classically forbidden region, namely the potential barriers, and
can be obtained from the Euclidean version of the action of eq. (\ref{2.7}) by Wick rotation 
$\tau = it, \beta = iT$.  The Euclidean propagator is
\begin{equation}
{\cal K}_E = \int{\cal D}\{\phi\} e^{-S_E}
\label{3.1}
\end{equation}
with the Euclidean action defined by
\begin{equation}
S_E = \int^{\tau_f}_{\tau_i} {\cal L}_E d\tau
\label{3.2}
\end{equation}
where the Euclidean Lagrangian is
\begin{equation}
{\cal L}_E = \frac {1}{2}m(\phi){\dot{\phi}}^2 + V(\phi)
\label{3.3}
\end{equation}
>From now on $\dot{\phi} = \frac {d\phi}{d\tau} $ denotes the imaginary time
derivative.  Since there are two kinds of barriers we expect two types of
instanton solutions.  The two classical solutions $\phi^{(i)}_c(\tau), i = 1,2 $,
which minimize the Euclidean action, eq. (\ref{3.2}), are found to be
instantons with nonvanishing topological charge, i.e.
\begin{eqnarray}
\tau &=& \frac{2\sqrt{a_+}}{\omega_0(a_- -2\sqrt{\lambda})\sqrt{a_-}}\nonumber\\
&&.\Bigg\{({\alpha_3}^2- {\alpha_1}^2)\Pi\left(\Phi^{(1)}(\phi^{(1)}_c(\tau)), {\alpha_3}^2, k\right)
+{\alpha_1}^2 F\left(\Phi^{(1)}(\phi^{(1)}_c(\tau)), k\right)\Bigg\}\qquad\
\label{3.4}
\end{eqnarray}
and
\begin{eqnarray}
\tau &=& \frac{2\sqrt{a_-}}{\omega_0(a_+ -2\sqrt{\lambda})\sqrt{a_+}}\nonumber\\
&&.\Bigg\{
({\alpha_4}^2 - {\alpha_2}^2)\Pi\left(\Phi^{(2)}(\phi^{(2)}_c(\tau)), {\alpha_4}^2, k\right)
+{\alpha_2}^2 F\left(\Phi^{(2)}(\phi^{(2)}_c(\tau)), k\right)\Bigg\}\qquad\
\label{3.5}
\end{eqnarray}
where $F$ denotes the incomplete elliptic intregral of the first kind
with elliptic modulus $k$, and 
$\Pi$ is Legendre's incomplete elliptic integral of the third kind
with the same elliptic modulus.  
The corresponding parameters are defined by
\begin{equation}
{\omega_0}^2=4K_1K_2s^2, \;\;\;\; k^2 =\frac{8\sqrt{\lambda(4+\lambda
\alpha^2)}}{a_+a_-},
\label{3.6a}
\end{equation}
where
\begin{equation}
{\alpha_1}^2=\frac{4\sqrt\lambda}{a_+},\;\; {\alpha_2}^2=\frac{4\sqrt\lambda}{a_-},\;\;
{\alpha_3}^2=\frac{2(a_- - 2\sqrt\lambda)}{(1-\alpha)a_+},\;\;
{\alpha_4}^2=\frac{2(a_+ - 2\sqrt\lambda)}{(1+\alpha)a_-}
\label{3.6b}
\end{equation}
and
\begin{equation}
a_+ = \sqrt{4+\lambda\alpha^2}+2\sqrt\lambda+\alpha\sqrt\lambda,\;\;
a_-=\sqrt{4+\lambda\alpha^2}+2\sqrt\lambda-\alpha\sqrt\lambda
\label{3.6c}
\end{equation}
Here $\Phi^{(i)}, i=1,2$, are functions of the classical trajectories $\phi^{(i)}_c(\tau)$, 
\begin{equation}
\Phi^{(1)}=\sin^{-1}\sqrt{\frac{(1-\sin\phi^{(1)}_c)a_+}
{2[a_+-2\sqrt\lambda(1+\sin\phi^{(1)}_c)]}}
\label{3.7}  
\end{equation}
\begin{equation}
\Phi^{(2)}=\sin^{-1}\sqrt{\frac{(1+\sin\phi^{(2)}_c)a_-}
{2[a_--2\sqrt\lambda(1-\sin\phi^{(2)}_c)]}}
\label{3.8}
\end{equation}
The trajectories of instantons $\phi^{(1)}_c $ and  $\phi^{(2)}_c $
are shown in Fig. 1.  At initial time $\tau_i = -\beta_1$ (or $\beta_2$)
the instanton $\phi^{(1)}_c $ (or $ \phi^{(2)}_c$)  starts from the
minimum of the potential well, for instance from $\phi_i = \sin^{-1}\alpha$
(or $\pi - \sin^{-1}\alpha$), and reaches the neighbouring minimum
$\phi_f = \pi - \sin^{-1}\alpha$ (or $2\pi - \sin^{-1}\alpha$)
at final time $\tau_f = \beta_1$ (or $\beta_2$), where
\begin{eqnarray}
\beta_1 &=& \frac{2\sqrt{a_+}}{\omega_0(a_- -2\sqrt{\lambda})\sqrt{a_-}}
\Bigg[({\alpha_3}^2
-{\alpha_1}^2)\Pi(\Phi^{(1)}(\alpha,\lambda), {\alpha_3}^2, k)\nonumber\\
&&\qquad\qquad\qquad\qquad\;\;\;
+{\alpha_1}^2F(\Phi^{(1)}(\alpha, \lambda), k)\Bigg],
\nonumber
\end{eqnarray}
\begin{equation}
\Phi^{(1)}(\alpha, \lambda)=\sin^{-1}\frac{1}{\alpha_3}
\label{3.9}
\end{equation}
and
\begin{eqnarray}
\beta_2 &=& \frac{2\sqrt{a_-}}{\omega_0(a_+ -2\sqrt{\lambda})\sqrt{a_+}}\Bigg[
({\alpha_4}^2-{\alpha_2}^2)\Pi(\Phi^{(2)}(\alpha,\lambda),{\alpha_4}^2,k)\nonumber\\
&&\qquad\qquad\qquad\qquad\;\;\;+{\alpha_2}^2F(\Phi^{(2)}(\alpha, \lambda), k)\Bigg],
\nonumber
\end{eqnarray}
\begin{equation}
\Phi^{(2)}(\alpha, \lambda)=\sin^{-1}\frac{1}{\alpha_4}
\label{3.10}
\end{equation}
It is interesting to observe that because of the position dependence of the mass
the instantons here do not need an infinite length of time to tunnel
through the barriers as in the usual instanton case of the double well and 
sine--Gordon potentials.  The minimum action, called instanton ``mass'',
evaluated along the instanton trajectories is
\begin{eqnarray}
S^{(1)}_c&=&\int^{\beta_2}_{-\beta_1}m({\dot{\phi}}^{(1)}_c)^2 d\tau =
4s\Bigg[\frac{4\sqrt\lambda-a_+}{\sqrt{a_+a_-}}\Pi(\Phi^{(1)}(\alpha,\lambda),
{\alpha_1}^2, k)\nonumber\\
&&\qquad\qquad +\frac{a_+-2\sqrt\lambda(1+\alpha)}{\sqrt{a_+a_-}}
F(\Phi^{(1)}(\alpha, k),k),\Bigg]
\label{3.11}
\end{eqnarray}
\begin{eqnarray}
S^{(2)}_c&=&\int^{\beta_2}_{-\beta_2}m({\dot{\phi}}^{(2)}_c)^2d\tau =
4s\Bigg[\frac{2\sqrt\lambda(a_++a_-)-a_+a_-}{\sqrt{a_+a_-^3}}\Pi(\Phi^{(2)}(\alpha,\lambda),
{\alpha_1}^2,k)\nonumber\\
&&\qquad\qquad +\frac{(a_--2\sqrt\lambda)a_++2\sqrt\lambda\alpha a_-}
{\sqrt{a_+a_-^3}}F(\Phi^{(2)}(\alpha, \lambda), k)\Bigg]
\label{3.12}
\end{eqnarray}
For the sake of simplicity we consider in the following the extremal model with
$\lambda << 1$.  This condition is of practical interest since the macroscopic
tunneling effect is negligible for the ferromagnetic particle with large spin $s$
except when $\lambda$ is extremely small \cite{12}.  The mass defined in eq.(\ref{2.6})
may then be replaced by a constant, i.e. $m\approx \frac{1}{2K_1}$.
With this replacement the effective Lagrangian eq. (\ref{3.3}) reduces to a simple model.
The classical solutions which minimise the Euclidean action are seen to be 
also instantons, i.e.
\begin{equation}
\phi^{(1)}_c(\tau)= 2\tan^{-1}\frac{1}{\alpha}\left(1+\sqrt{1-\alpha^2}
{\frac{e^{\sqrt{1-\alpha^2}\omega_0(\tau -\tau_0)}+f^{(1)}(\alpha)}
{e^{\sqrt{1-\alpha^2}\omega_0(\tau -\tau_0)}-f^{(1)}(\alpha)}}\right),
\label{3.13}
\end{equation}
\begin{equation}
\phi^{(2)}_c(\tau) =2\tan^{-1}\frac{1}{\alpha}\left(1 +\sqrt{1-\alpha^2}
{\frac {e^{-\sqrt{1-\alpha^2}\omega_0(\tau-\tau_0)}+f^{(2)}(\alpha)}
{e^{-\sqrt{1-\alpha^2}\omega_0(\tau-\tau_0)}-f^{(2)}(\alpha)}}\right)
\label{3.14}
\end{equation}
where
\begin{equation}
f^{(1)}(\alpha)=\frac{1-\alpha+\sqrt{1-\alpha^2}}{1-\alpha-\sqrt{1-\alpha^2}},\;\;\;
f^{(2)}(\alpha)=\frac{1+\alpha+\sqrt{1-\alpha^2}}{1+\alpha-\sqrt{1-\alpha^2}}
\label{3.15}
\end{equation}
These solutions of the simplified model also coincide with the 
$\lambda\rightarrow 0$ ($k\rightarrow 0$) limit of 
eqs.(\ref{3.4},\ref{3.5}).

The trajectories of instantons $\phi^{(i)}_c, i=1, 2, $ are the same as those
shown in Fig. 1, but with $\beta_1$ and $\beta_2$ tending to infinity.  $\tau_0$ is the
``position'' of the instanton.  When $\tau = \tau_0$ the instantons arrive at the
centres of their barriers ($\phi = \frac{\pi}{2}$, mod $2\pi$,
for $\phi^{(1)}_c$ and $\phi = \frac{3\pi}{2}, $mod$2\pi$, for $\phi^{(2)}_c$).
The minimum actions (masses) evaluated along the trajectories of the
instantons are seen to be 
\begin{equation}
S^{(1)}_c=\int^{\infty}_{-\infty}m ({\dot{\phi}}^{(1)}_c)^2 d\tau
= s\sqrt\lambda(2\sqrt{1-\alpha^2}-\alpha\pi +2\alpha\sin^{-1}\alpha),
\label{3.16}
\end{equation}
\begin{equation}
S^{(2)}_c=\int^{\infty}_{-\infty}m  ({\dot{\phi}}^{(2)}_c)^2 d\tau
=s\sqrt\lambda(2\sqrt{1-\alpha^2}+\alpha\pi +2\alpha\sin^{-1}\alpha)
\label{3.17}
\end{equation}
The amplitudes for tunneling through the barriers of type ``1'' (those of small
size with centres at $\phi = \frac{\pi}{2}$, mod $2\pi$) and
of type ``2'' (those of large size with centres at $\phi = \frac{3\pi}{2} $, mod $2\pi$)
can be naively calculated by evaluating the Feynman propagator eq. ({2.7}).  However,
it is not clear what the physical result relating to the two tunneling propagators
would be.  In the double well potential case the Feynman propagator
is related to the level splitting of the degenerate ground states.  In our case we would
have two different level splittings $\triangle\epsilon_1$
and $\triangle\epsilon_2$.  To solve this apparent paradox we calculate in the following
section the energy spectrum with the help of Bloch theory for the periodic
potential.  \newline\\

\noindent
{\large \bf 4.  The energy spectrum}\newline\\

We let $|\epsilon_0, \Phi^{(1)}_{2n}>$ be the ground state of the harmonic oscillator approximated
Hamiltonian ${\hat H}^{(1)}_0$ in the $2n$th well around the local
potential minimum at $\Phi^{(1)}_{2n} = 2n\pi + \sin^{-1}\alpha $
such that
\begin{equation}
{\hat{H}}^{(1)}_0|\epsilon_0, \Phi^{(1)}_{2n}> = \epsilon_0|\epsilon_0, \Phi^{(1)}_{2n}>
\label{4.1}
\end{equation}
where the Hamiltonian is seen to be 
\begin{equation}
{\hat H}^{(1)}_0 = \frac{ {\hat p}^2}{2m} + \frac{1}{2}m{\omega_0}^2
(\phi - \Phi^{(1)}_{2n})^2
\label{4.2}
\end{equation}
and $\phi - \Phi^{(1)}_{2n} $ is the angle deviation from $\Phi^{(1)}_{2n}$.  Similarly
$|\epsilon_0, \Phi^{(2)}_{2n+1}>$ denotes the ground state of the harmonic 
oscillator approximated Hamiltonian ${\hat{H}}^{(2)}_0$ in the $(2n+1)$th well around the local
potential minimum $\Phi^{(2)}_{2n+1} = (2n+1)\pi-\sin^{-1}\alpha$.  ${\hat{H}}^{(2)}_0$
is seen to be the same as ${\hat {H}}^{(1)}_0$ but with the angle $\phi - \Phi^{(2)}_{2n+1}$
understood as
the deviation from the equilibrium position $\Phi^{(2)}_{2n+1}$.  According
to Bloch theory the true ground state of the system is a superposition of
$|\epsilon_0, \Phi^{(i)}>, i = 1,2$, i.e.
\begin{equation}
|\psi> = \sum_n\left(e^{i\xi\Phi^{(1)}_{2n}}|\epsilon_0, \Phi^{(1)}_{2n}>
+ e^{i\xi\Phi^{(2)}_{2n+1}}|\epsilon_0, \Phi^{(2)}_{2n+1}>\right)
\label{4.3}
\end{equation}
which is a Bloch state.  $\xi$ denotes the Bloch wave vector and should be an integer in
our case because of the $2\pi$ boundary condition \cite{13}.  We then have
\begin{equation}
\hat{H}|\psi> = E |\psi>
\label {4.4}
\end{equation}
$\hat{H}$ is the Hamiltonian of the system with periodic potential and $E$ is the true ground state
energy to be determined.  Simple calculation leads to (taking into account 
only nearest neighbours and remembering that as in eq.(\ref{2.2}) the spin 
coherent state representation leads to wave functions with topological phase, 
e.~g.~ $\langle\phi|\epsilon_0,\Phi^{(1)}_{2n}\rangle = 
u_0(\phi-\Phi^{(1)}_{2n})e^{i(\phi-\Phi^{(1)}_{2n})s}$)
\begin{equation}
E = \epsilon_0 + \triangle\epsilon_1\cos\left[(s+\xi)(\pi-2\sin^{-1}\alpha)\right]
+\triangle\epsilon_2\cos\left[(s+\xi)(\pi +2\sin^{-1}\alpha)\right]
\label{4.5}
\end{equation}
where
\begin{equation}
\triangle\epsilon_1 = -\int u_0(\phi -\Phi^{(1)}_{2n})\hat{H}u_0(\phi-\Phi^{(2)}_{2n+1})d\phi
\label{4.6}
\end{equation}
\begin{equation}
\triangle\epsilon_2= -\int u_0(\phi-\Phi^{(1)}_{2n})\hat{H}u_0(\phi-\Phi^{(2)}_{2n-1})d\phi
\label{4.7}
\end{equation}
are just the level shifts due to tunneling through the small (type 1) and
larger (type 2) barriers respectively and they are the same for all $n$.  
The function $u_0(\phi -\Phi^{(i)}) $ denotes the ground state wave function of Hamiltonian
$\hat{H}^{(i)}_0$.  The spin number $s$ in eq. (\ref {4.5}) comes from the inner
product of the spin--coherent--states like the phase factor in eq.(\ref{2.2}).  
The expressions $\triangle\epsilon_1$ and $\triangle\epsilon_2$ can be evaluated
with standard instanton methods, as we shall show in the following.   \newline\\

\newpage
\noindent
{\large \bf 5. Amplitude for tunneling through the asymmetric twin barrier}\newline\\

We start from the Euclidean version (imaginary time) of the element of the evolution
operator, eq.(\ref{2.2}), that is,
\begin{equation}
<{\bf n}_f|e^{-2\beta\hat{H}}|{\bf n}_i> = e^{-i(\phi_f-\phi_i)s}
{\cal K}(\phi_f, \tau_f; \phi_i, \tau_i)
\label{5.1}
\end{equation}
Inserting the complete set of eigenstates of the harmonic oscillator approximated
Hamiltonians $\hat{H}^{(1)}_0$ and $\hat{H}^{(2)}_0$ on the left hand side of
eq.(\ref{5.1}) and taking the large time ($\beta$) limit, we obtain the 
relations between tunneling propagators and the level shifts $\triangle\epsilon_1$,
$\triangle\epsilon_2$ respectively, where $\epsilon_0$
is the corresponding ground state energy for both ${\hat{H}}^{(1)}_0$
and ${\hat{H}}^{(2)}_0$, i.e.
\begin{eqnarray}
& &u_0\left(\Phi^{(1)}_{2n}\right)u_0\left(\Phi^{(2)}_{2n+1}\right)
e^{-2\beta\epsilon_0}\sinh(2\beta\triangle\epsilon_1)
\nonumber\\& & \qquad\quad = e^{-is(\pi-2\sin^{-1}\alpha)}{\cal K}^{(1)}\left(\phi_f=\Phi^{(2)}_{2n+1}, 
\beta; \phi_i = \Phi^{(1)}_{2n}, -\beta\right)\qquad
\label{5.2}
\end{eqnarray}
and
\begin{eqnarray}
& & u_0\left(\Phi^{(2)}_{2n-1}\right)u_0\left(\Phi^{(1)}_{2n}\right)e^{-2\beta\epsilon_0}
\sinh(2\beta\triangle\epsilon_2)\nonumber\\
& & \qquad\quad = e^{-is(\pi+2\sin^{-1}\alpha)}{\cal K}^{(2)}\left(\phi_f=\Phi^{(1)}_{2n}, \beta; 
\phi_i = \Phi^{(2)}_{2n-1}, -\beta\right)\qquad
\label{5.3}
\end{eqnarray}
In the following we evaluate the Feynman propagators about the instanton
trajectories $\phi^{(1)}_c$ and $\phi^{(2)}_c$ and compare the results
with eqs.(\ref{5.2}) and (\ref{5.3}) to determine the level shifts
$\triangle\epsilon_1, \triangle\epsilon_2$.  The propagators ${\cal K}^{(i)}, i =1,2, $
are defined by eqs.(\ref{3.1}) to (\ref{3.3}).  We expand $\phi$ about the classical
trajectories $\phi^{(i)}_c$, 
\begin{equation}
\phi = \phi^{(i)}_c + \eta^{(i)}
\label{5.4}
\end{equation}
where $\eta^{(i)}$ denotes the small fluctuation about the classical trajectory
$\phi^{(i)}_c$ with fixed end points such that
\begin{equation}
\eta^{(i)}(\phi_f) = \eta^{(i)}(\phi_i) = 0
\label{5.5}
\end{equation}
Substituting eq.(\ref{5.4}) into the propagator eq.(\ref{3.1}) and keeping terms up to
and including those quadratic in $\eta^{(i)}$ for the one--loop approximation, one has
\begin{equation}
{\cal K}^{(i)} = e^{-S^{(i)}_c} I^{(i)}
\label{5.6}
\end{equation}
where
\begin{equation}
I^{(i)} = \int^{\eta^{(i)}(\phi_f)=0}_{\eta^{(i)}(\phi_i)= 0} {\cal D}\{\eta^{(i)}\}e^{-\triangle S^{(i)}}
\label{5.7}
\end{equation}
is the fluctuation functional integral with fluctuation action
\begin{equation}
\triangle S^{(i)} = \int \eta^{(i)}\hat{M}^{(i)}\eta^{(i)} d\tau
\label{5.8}
\end{equation}
where
\begin{equation}
\hat{M}^{(i)} = \frac{1}{2}\left[-m\frac{d^2}{d\tau^2} + V^{\prime\prime}(\phi^{(i)}_c)\right]
\label{5.9}
\end{equation}
is the operator of the second variation of the action, and
$$ V^{\prime\prime}(\phi^{(i)}_c) = \frac {\partial^2V(\phi)}{\partial\phi^2}\Bigg|_{\phi = \phi^{(i)}_c}.
$$
We let $\{\psi^{(i)}_n\}$ be the set of normalized eigenfunctions of the operator
${\hat{M}}^{(i)}$ with eigenvalues $\{E^{(i)}_n\}$ such that
\begin{equation}
{\hat{M}}^{(i)}\psi^{(i)}_n = E^{(i)}_n\psi^{(i)}_n
\label{5.10}
\end{equation}
We may expand $\eta^{(i)}$ in terms of $\psi^{(i)}_n$, 
\begin{equation}
\eta^{(i)} = \sum_nc^{(i)}_n\psi^{(i)}_n
\label{5.11}
\end{equation}
Then the functional integral can be formally expressed as
\begin{equation}
I^{(i)}=\left| \frac{\partial\eta^{(i)}}{\partial c^{(i)}_n}\right| \prod_{n=0}(\frac{\pi}{E^{(i)}_n})^{\frac{1}{2}}
\label{5.12}
\end{equation}
The first factor on the right is the Jacobi determinant of the transformation (\ref{5.11}).
Because of the time translation symmetry of the action there exists naturally a zero mode
of $\hat{M}^{(i)}$ (with $E^{(i)}_0 =0$).  This zero mode is seen to be just the time derivative of
the classical  solution, i.e.
\begin{equation}
\psi^{(i)}_0 = N^{(i)}\frac{d\phi^{(i)}_c}{d\tau}
\label{5.13}
\end{equation}
where $N^{(i)}$ is a normalization constant.  The fluctuation functional is divergent,
as seen from eq.(\ref{5.12}).  This well known problem of the quantization 
about a classical solution can be cured by the Faddeev--Popov (FP) technique \cite{14}
or in a more systematic way using the BRST transformation\cite{15}.  The result is
that the integration over the zero mode variable, $dc_0$, is replaced by an integration over the
collective coordinate, i.e. the instanton position $d\tau_0$ which leads
to a factor $2\beta$ and an FP determinant \cite{15} $\sqrt{D}$ given by
\begin{equation}
D = \int^{\infty}_{-\infty}({\dot{\phi}}^{(i)}_c)^2 d\tau = 2K_1S^{(i)}_c
\label{5.14}
\end{equation}
One then has
\begin{equation}
I^{(i)} = 2\beta\sqrt{2K_1S^{(i)}_c} I^{(i)}_0
\label{5.15}
\end{equation}
where
\begin{equation}
I^{(i)}_0 = \left|\frac{\partial\eta^{(i)}}{\partial c^{(i)}_n}\right|\prod_{n\neq 0}(\frac{\pi}{E^{(i)}_n})
^{\frac{1}{2}}
\label{5.16}
\end{equation}
which is not easy to evaluate directly although it is divergence free.  We now evaluate $I^{(i)}$ in
an alternative way known as the shift method \cite{16} in path integral theory.
Introducing the transformation
\begin{equation}
\eta^{(i)}(\tau)=y^{(i)}(\tau)+\dot{\phi}^{(i)}_c(\tau)
\int^{\tau}_{-\beta}\frac{{\ddot{\phi}}^{(i)}_c(\tau^{\prime)}}{{\dot{\phi}}^{(i)}_c(\tau^{\prime})}y^{(i)}
(\tau^{\prime})d\tau^{\prime}
\label{5.17}
\end{equation}
which converts the functional integral $I^{(i)}$ into a
Gaussian integral.  We obtain 
\begin{equation}
I^{(i)}=(\frac{1}{2\pi})^{\frac{1}{2}}\left[{\dot{\phi}}^{(i)}_c(\beta){\dot{\phi}}^{(i)}_c(-\beta)\right]
^{-\frac{1}{2}}\left[\int^{\beta}_{-\beta}\frac {d\tau}{m({\dot{\phi}}^{(i)}_c)^2}\right]^{-\frac{1}{2}}
\label{5.18}
\end{equation}
Substituting the explicit formula for ${\dot{\phi}}^{(i)}_c$ which is the time derivative of
eqs.(\ref{3.13}), (\ref{3.14}), we find the functional fluctuation integrals are the same
for both instantons $\phi^{(1)}_c$ and $\phi^{(2)}_c$ in the large time limit, i.e. 
$\beta\rightarrow\infty$, so that
\begin{equation}
I^{(1)}= I^{(2)} = \frac{1}{\sqrt{2\pi}} (1-\alpha^2)^{\frac{1}{4}}\lambda^{\frac{1}{4}}s^{\frac{1}{2}}
\label{5.19}
\end{equation}
To obtain the desired result the fluctuation functional integral expression may be
reexpressed by comparison of eqs. (\ref{5.15}), (\ref{5.16}) and (\ref{5.19}) as
\begin{equation}
I^{(i)} = 2\beta\sqrt{2K_1S^{(i)}_c}(\frac{E^{(i)}_0}{\pi})^{\frac{1}{2}}
\frac{1}{\sqrt{2\pi}}(1-\alpha^2)^{\frac{1}{4}}\lambda^{\frac{1}{4}}s^{\frac{1}{2}}
\label{5.20}
\end{equation}
This expression involves explicitly the zero mode eigenvalue $E^{(i)}_0$ which
vanishes when $\beta\rightarrow\infty$.  We can find the explicit time dependence
of $E^{(i)}_0(\beta)$ by using a socalled boundary perturbation method \cite{1,2}
which leads to the formula
\begin{equation}
m\left[{\dot{\phi}}^{(i)}_c(\beta){\ddot{\phi}}^{(i)}_c(\beta)
-{\dot{\phi}}^{(i)}_c(-\beta){\ddot{\phi}}^{(i)}_c(-\beta)\right]
=-2E^{(i)}_0\int^{\beta}_{-\beta}({\dot{\phi}}^{(i)}_c)^2d\tau
\label{5.21}
\end{equation}
where 
$$
{\dot{\phi}}^{(i)}_c(\beta)\equiv\frac{d{\phi}^{(i)}_c(\tau)}{d\tau}\Bigg|_{\tau = \beta},\;\;\;\;
{\ddot{\phi}}^{(i)}_c(\beta)\equiv\frac{d^2\phi^{(i)}_c(\tau)}{d\tau^2}\Bigg|_{\tau=\beta}
$$
After evaluation of both sides of eq.(\ref{5.21}) we again take the large time limit.  The
time dependence of the zero mode eigenvalue is then found to be
\begin{equation}
E^{(i)}_0=\frac{4\lambda(1-\alpha^2)^{\frac{5}{2}}s^2\omega_0}{S^{(i)}_c}
e^{-2\sqrt{1-\alpha^2}\omega_0\beta}
\label{5.22}
\end{equation}
Substituting this expression into eq.(\ref{5.20}) the desired result of the Feynman
propagator due to tunneling of one instanton is found to be
\begin{equation}
K^{(i)}_{(1)}=2\beta\frac{\sqrt 2}{\pi}(1-\alpha^2)^{\frac{3}{2}}\lambda^{\frac{1}{2}}s\omega_0
e^{-\sqrt{1-\alpha^2}\omega_0\beta} e^{-S^{(i)}_c}
\label{5.23}
\end{equation}
The subscript $(1)$ denotes the one instanton contribution to the propagator.  

The path integral quantization requires a sum over all possible paths.  In the present case
the paths fall into an infinite number of classes consisting of one instanton plus
an arbitrary number of instanton--anti--instanton pairs for a given time interval, i.e.
\begin{equation}
{\cal K}^{(i)} =\sum^{\infty}_{n=0} K^{(i)}_{(2n+1)} (\phi_f, \beta;\phi_i, -\beta)
\label{5.24}
\end{equation}
where $K^{(i)}_{(2n+1)}$ is the Feynman kernel of the contribution of one instanton
plus $n$ pairs.  The propagator with $n$ pairs is found to be
\begin{eqnarray}
K^{(i)}_{(2n+1)}&=&\frac{(2\beta)^{2n+1}}{(2n+1)!}\left[\frac{2}{\pi^{\frac{1}{2}}}
(1-\alpha^2)^{\frac{5}{4}}s^{\frac{1}{2}}\omega_0\lambda^{\frac{1}{4}}\right]^{2n+1}
\frac {\lambda^{\frac{1}{4}}(1-\alpha^2)^{\frac{1}{4}}}{\sqrt{2\pi}}.\nonumber\\
&&e^{-\sqrt{1-\alpha^2}\omega_0\beta}e^{-(2n+1)S^{(i)}_c}
\label{5.25}
\end{eqnarray}
The final result of the tunneling propagator is
\begin{equation}
{\cal K}^{(i)} =\frac{\lambda^{\frac{1}{4}}(1-\alpha^2)^{\frac{1}{4}}}{\sqrt{2\pi}}
e^{-\sqrt{1-\alpha^2}\omega_0\beta}\sinh \left[2\beta\frac{2}{\sqrt{\pi}}
(1-\alpha^2)^{\frac{5}{4}}s^{\frac{1}{2}}\omega_0\lambda^{\frac{1}{4}}
e^{-S^{(i)}_c}\right]
\label{5.26}
\end{equation}
We have seen that the quantum fluctuation part of the tunneling propagator is the same
for both types of instantons $\phi^{(1)}_c$ and $\phi^{(2)}_c$.  The only
difference comes from the classical action.  Comparing eq.(\ref{5.26}) with eqs.(\ref{5.2})
and (\ref{5.3}) we obtain the corresponding level shifts $\triangle\epsilon_i, i=1,2,$
\begin{equation}
\triangle\epsilon_i = P(\alpha, s, \lambda) e^{-S^{(i)}_c}
\label{5.27}
\end{equation}
The prefactor is given by
\begin{equation}
P(\alpha, s, \lambda) = \frac{2}{\sqrt\pi}(1-\alpha^2)^{\frac{5}{4}}s^{\frac{1}{2}}\omega_0
\lambda^{\frac{1}{4}}
\label{5.28}
\end{equation}
\newline\\

\noindent
{\large \bf 6.  Quenching of macroscopic quantum coherence}\newline\\

The small ferromagnetic particle described by the Hamiltonian of eq.(\ref{2.1}) in the
limit of large spin $s$ has two degenerate easy directions in the absence
of an external magnetic field, corresponding to the two degenerate
ground states at $\phi = 0, \pi$.  Quantum tunneling mixes the degenerate ground
states and the total moment of the small ferromagnetic particle
may resonate between the degenerate easy directions, a phenomenon 
known as macroscopic quantum coherence (MQC).  It has been pointed
out that the MQC is quenched for half integer spin $s$, and that the 
quenching of MQC is seen as a consequence of Kramer's degeneracy \cite{17}.  
In earlier work \cite{10} we demonstrated that the quenching of MQC should be
understood as the quenching of an energy band, and is due to the destructive
interference of topological phases of the tunneling amplitude in eq.(\ref{2.2}).
In the present case with a small applied magnetic field the easy
directions are shifted to $\phi = \sin^{-1}\alpha$ and $\pi - \sin^{-1}\alpha$.
The quantum tunneling takes place either through the small barrier
with angle difference $\pi - 2\sin^{-1}\alpha $ or through the larger
barrier with angle difference $\pi +2\sin^{-1}\alpha$.  The total magnetic
moment of the small ferromagnetic particle may also resonate between the
degenerate easy directions.  It may be interesting to see the effect of the
topological phase on MQC.  To this end we rewrite the energy
spectrum of eq.(\ref{4.5}) as
\begin{eqnarray}
E&=&\epsilon_0+(\triangle\epsilon_1+\triangle\epsilon_2)\cos\left[(\xi+s)\pi\right]
\cos\left[2(\xi+s)\sin^{-1}\alpha\right]\nonumber\\
&&-(\triangle\epsilon_1 - \triangle\epsilon_2)
\sin\left[(\xi+s)\pi\right]\sin\left[2(\xi+s)\sin^{-1}\alpha\right]
\label{6.1}
\end{eqnarray}
For a weak magnetic field $\alpha <<1$ .  Remembering that $\xi$ can assume only
either of the two values $0$ and $1$ in the first Brillouin zone, the energy
spectrum can be rewritten in terms of the explicit level shifts of eqs.(\ref{5.27})
and the classical actions of eqs.(\ref{3.16}) and (\ref{3.17}) as  
\begin{eqnarray}
E&=&\epsilon_0+2P(\alpha,s,\lambda)e^{-2s\sqrt\lambda}
\Bigg\{\cos[(s+\xi)\pi]\cos(2s\alpha)\cosh(s\sqrt\lambda\alpha\pi)\nonumber\\        
&&+\sin[(s+\xi)\pi]\sin(2s\alpha)\sinh(s\sqrt\lambda\alpha\pi)\Bigg\}
\label{6.2}
\end{eqnarray}
In view of the second term the energy band cannot be quenched for half integer
spin $s$, since Kramer's degeneracy is removed by an external magnetic field
except when $s\sqrt\lambda\alpha\pi $ is negligibly small. There is, in fact, another type
of quenching of MQC here for both integer spin $s$ if
\begin{equation}
2\alpha s = \left(n+\frac12\right) \pi
\label{6.3}
\end{equation}
and half integer spin $s$ if
\begin{equation}
2\alpha s = n \pi
\label{6.4}
\end{equation}
where $n$ is an integer.  This quenching of MQC is irrelevant to Kramer's degeneracy.
\newline\\

\noindent
{\large \bf 7. Conclusions}\newline\\
In the above we presented a first report on the application of the instanton method
to the analysis of quantum tunneling in a periodic potential with asymmetric twin
barriers.  There are two types of instanton solutions corresponding to the
two barriers of different sizes.  The instanton method alone can only yield the separate
level shifts due to tunneling.  The energy spectrum, eq.(\ref{4.5}), and finally
eqs.(\ref{6.2}) and (\ref{6.3}), here first derived and calculated in the one--loop 
approximation, were obtained with the help of Bloch theory and involve
both level shifts.  A new kind of quenching of MQC is also discussed.
\newline\\

\noindent
{\large \bf Acknowledgements}\newline\\
J.--Q. Liang acknowledges support of the Deutsche Forschungsgemeinschaft and
J.--G. Zhou support of the A.von Humboldt Foundation.  J.--Q. Liang and
F.--C. Pu also acknowledge support of the National Natural Science Foundation
of China.\newline\\

\newpage
\begin{center}
{\large \bf Figure Caption}
\end{center}
The periodic potential with asymmetric twin barriers and the instanton trajectories 


\begin{thebibliography}{99}
\bibitem{1} E. Gildener and A. Patrascioiu, Phys. Rev. {\bf D16}, 423 (1977).
\bibitem{2} J.--Q. Liang and H. J. W. M\"uller--Kirsten, Phys. Rev. {\bf D45}, 2963 (1992) and
{\bf 48}, 964(E)(1993); S.K.Bose and H. J. W. M\"uller--Kirsten, Phys.Lett. {\bf A162},79 (1992).
\bibitem{3}J. Zinn--Justin, ``Quantum Field Theory and Critical Phenomena'' (Oxford
Science Publication, 1984).
\bibitem{4} N. S. Manton and T, S. Samols, Phys. Lett. {\bf B207}, 179 (1988).
\bibitem{5} J.--Q. Liang, H. J. W, M\"uller--Kirsten and D. H. Tchrakian, Phys. Lett. {\bf B282},
105(1992).
\bibitem{6} J.--Q. Liang and H. J. W. M\"uller--Kirsten, Phys. Rev. {\bf D46}, 4685(1992),
{\bf D50}, 6519(1994), {\bf D51}, 718(1995). For a review see J.--Q. Liang and H. J. W.
M\"uller--Kirsten, ``Topics in Quantum Field Theory -- Modern Methods in Fundamental Physics'',
ed. D. H. Tchrakian (World Scientific, 1996), pp. 54 --68.
\bibitem{7} A. J. Legget, S. Chakravarty, A. T. Dorsey, M. P. A. Fisher, A. Garg and W. 
Zwerger, Rev. Mod. Phys. {\bf 59}, 1 (1987).
\bibitem{8} ``Quantum Tunneling of Magnetization--QTM'94'', ed. L. Gunther and
B. Barbara, NATO ASI Series, Vol. {\bf 301} (Kluver Acad. , 1995).
\bibitem{9}  E. Fradkin, ``Field Theories of Condensed Matter Systems'',(Addison--Wesley, 1991).
\bibitem{10} J.--Q. Liang, H. J. W. M\"uller--Kirsten and Jian--Ge Zhou, Report KL-TH 96/6
and references therein.
\bibitem{11} E.S. Abers and B.W. Lee, Phys. Rep.{\bf 9},1 (1973).
\bibitem{12} J.--Q. Liang, H.J. W. M\"uller--Kirsten, Jian--Ge Zhou and F.--C. Pu, Report
KL--TH 96/7.
\bibitem{13} The Bloch wave vector $\xi$ is determined by the $2\pi$ periodicity of the
Bloch phase factor, i.e. $e^{i\xi\Phi^{(1)}_{2n}}= e^{i\xi\phi^{(1)}_{2(n+1)}}$
or equivalently $e^{i\xi\Phi^{(2)}_{2(n+1)+1}} = e^{i\xi\Phi^{(2)}_{2n+1}}$.
The natural length of our super--latice is $2\pi$.  The one--dimensional periodic
potential is generated by successive $2\pi$ extensions.  
\bibitem{14} L.D. Faddeev and V. N. Popov, Phys. Lett. {\bf B25},29 (1967).
\bibitem{15} Jian--Ge Zhou, F. Zimmerschied, J.--Q. Liang and H. J. W. M\"uller--Kirsten,
Phys. Lett. {\bf B365}, 163 (1996).
\bibitem{16} R. F. Dashen, B. Hasslacher and A. Neveu, Phys. Rev. {\bf D10}, 4114 (1974).
\bibitem{17} Kramer's theorem says that for half integer spin $s$ the degeneracy
of the ground state cannot be removed in the absence of an external magnetic field.  
\end{thebibliography}
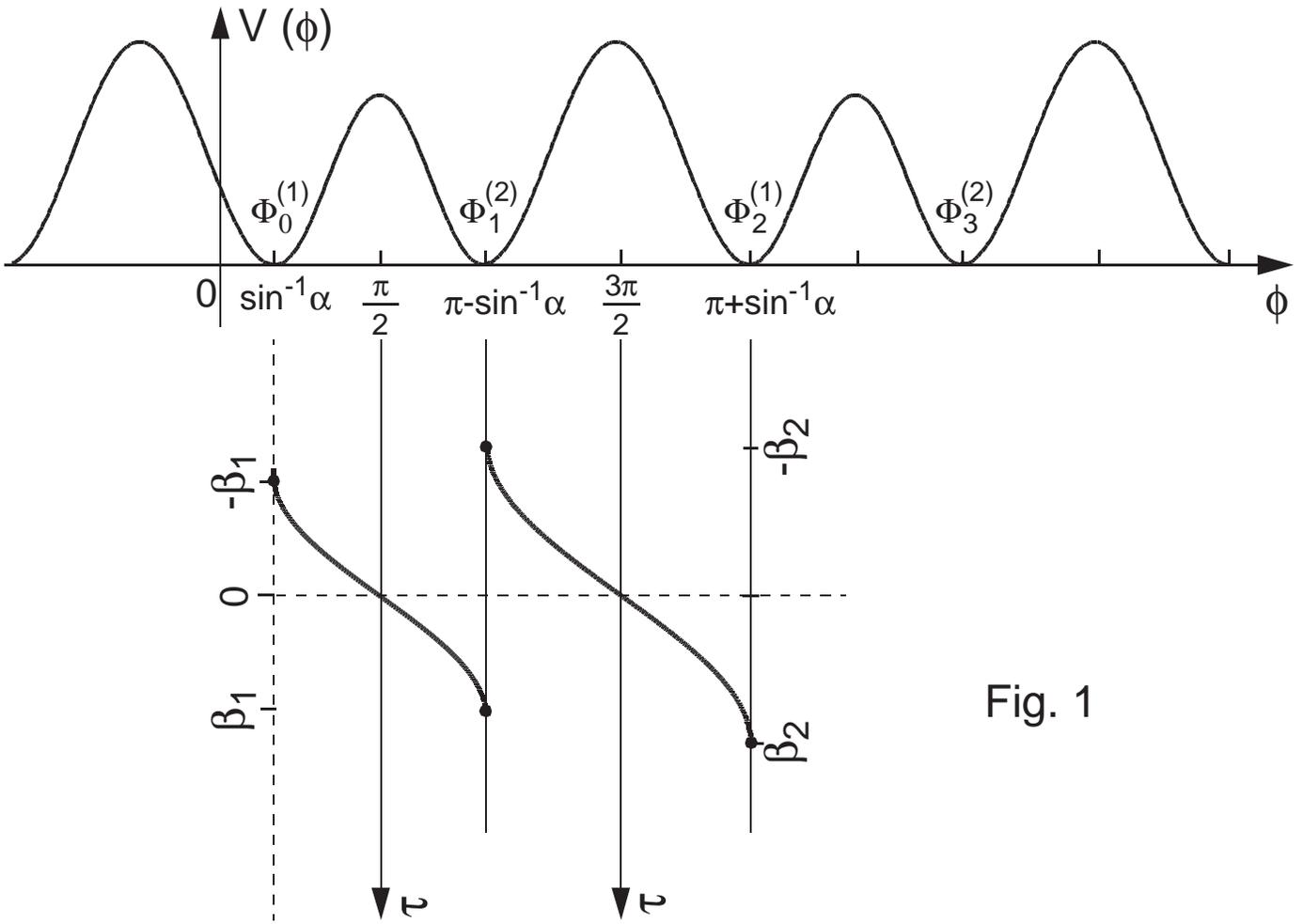
\end{document}